\documentclass{appolb}
\usepackage{epsfig}
\usepackage{cite}
\usepackage{amsmath}

\begin{document}
\pagestyle{plain}
\title{Kaon Production and Interaction
\thanks{Presented at MESON 2002}
}
\author{M. Wolke for the COSY--11 Collaboration
\address{Institut f\"ur Kernphysik, Forschungszentrum J\"ulich, 
D--52425 J\"ulich, Germany}
}
\maketitle
\begin{abstract}
Exclusive data on both the elementary kaon and antikaon production 
channels have been taken at the cooler synchrotron COSY in 
proton--proton scattering.

In the kaon--hyperon production an enhancement by one order of 
magnitude of the $\Lambda/\Sigma^0$ ratio has been observed at excess 
energies below $\mbox{Q} = 13\,\mbox{MeV}$ compared to data at higher 
excess energies ($\mbox{Q} \ge 300\,\mbox{MeV}$).
New results obtained at the COSY--11 facility explore the transition 
region between the regime of this low--energy $\Sigma^0$ suppression 
and excess energies of $60\,\mbox{MeV}$.
A comparison of the energy dependence of the $\Lambda$ and $\Sigma^0$ 
total cross sections exhibits distinct qualitative differences 
between both hyperon production channels. 

Studies of kaon--antikaon production have been motivated especially 
by the ongoing discussion about the nature of the scalar resonances 
$f_0(980)$ and $a_0(980)$ coupling to the $K \overline{K}$ channel.
For the reaction $pp \rightarrow pp K^+ K^-$ a first total cross 
section value is reported at an excess energy of $\mbox{Q} = 
17\,\mbox{MeV}$, i.e.\ below the $\phi$ threshold. 
Calculations obtained within an OBE model indicate that the energy 
dependence of the available total cross section data close to 
threshold is rather difficult to reconcile with the assumption of a 
phase--space behaviour modified predominantly by the proton--proton 
final state interaction.
\end{abstract}

\PACS{13.75.-n, 14.20.Jn, 14.40.Aq, 25.40.Ep}
  
\section{Introduction}
In elementary hadronic interactions with no strange valence quark in 
the initial state the associated strangeness production provides a 
powerful tool to study reaction dynamics by introducing an 
``impurity'' to hadronic matter. 
Thus, quark model concepts might be related to mesonic or baryonic 
degrees of freedom, with the onset of quark degrees of freedom 
expected for kinematical situations with large enough transverse 
momentum transfer.

In close--to--threshold production experiments, effects of low energy 
scattering are inherent to the observables due to strong final state 
interactions.
Consequently, the data allow to constrain microscopic interaction 
models especially in the case where direct scattering experiments are 
difficult to perform.

\section{Exclusive Kaon--Hyperon Experiments in Proton--Proton 
Collisions}
Exclusive data on $\Lambda$ and $\Sigma^0$ production in 
proton--proton scattering have been taken at the COSY--11 facility 
at equal excess energies close to threshold~\cite{bal98,sew99}.
In the energy range up to $\mbox{Q} = 13\,\mbox{MeV}$ the energy 
dependence is better described by a phase space behaviour modified by 
the proton--hyperon final state interaction (FSI) than by a pure 
phase space behaviour~\cite{sew99}.
However, the most striking feature of the data is the observed 
$\Sigma^0$ suppression with 
\begin{equation}
\label{eq_lsratio}
{\cal{R}}_{\Lambda/\Sigma} \left(\mbox{Q} \le 13\,\mbox{MeV}\right)= 
  \frac{\sigma\left(pp \rightarrow p K^+ \Lambda \right)}
       {\sigma\left(pp \rightarrow p K^+ \Sigma^0 \right)} = 
  28^{+6}_{-9}\, ,
\end{equation}
while at excess energies $\ge 300\,\mbox{MeV}$ this ratio is known to 
be about 2.5~\cite{bal88}.
Considering only $\pi$ exchange, data on $\pi$ induced hyperon 
production via $\pi N \rightarrow K \Lambda \left(\Sigma^0\right)$ 
result in a ratio of ${\cal{R}}_{\Lambda/\Sigma} \approx 
0.9$~\cite{gas00}, clearly underestimating the experimental value 
of~\eqref{eq_lsratio}.
Kaon exchange essentially relates the ratio 
${\cal{R}}_{\Lambda/\Sigma}$ to the ratio of coupling constants 
squared $\mbox{g}^2_{N \Lambda K}/\mbox{g}^2_{N \Sigma K}$.
Although there is some uncertainty in the literature on their values, 
a $\Lambda/\Sigma^0$ production ratio of 27 follows from the suitable 
choice of the SU(6) prediction~\cite{gnyk}, in good agreement with the 
experiment.
However, effects of final state interaction as well as the importance 
of $\pi$ exchange for $\Sigma^0$ production are completely neglected 
by this simple estimate.

Inclusive $K^+$ production data in $pp$ scattering taken at the 
SPES~4 facility at SATURNE at an excess energy of $252\,\mbox{MeV}$ 
with respect to the $p K^+ \Lambda$ threshold show enhancements in 
the invariant mass distribution at the $\Lambda p$ and $\Sigma N$ 
thresholds of similar magnitude~\cite{sie94}.
With only the $K^+$ being detected, it is not clear whether the 
enhancement at the $\Sigma N$ threshold originates from $\Sigma$ 
production.
Qualitatively, a strong $\Sigma N \rightarrow \Lambda p$ final state 
conversion might account for both the inclusive SPES~4 results as well 
as the $\Sigma^0$ depletion in the COSY--11 data.
Evidence for $\Sigma N \rightarrow \Lambda p$ conversion effects is 
known from exclusive hyperon data via $K^- d \rightarrow 
\pi^- \Lambda p$~\cite{kminusd}, and from hypernuclear physics, with 
$\Sigma N \rightarrow \Lambda N$ as the dominating decay channel of 
$\Sigma$--hypernuclei~\cite{ose90}.

In exploratory calculations performed within the framework of the 
J\"ulich meson exchange model~\cite{gas00}, taking into account both 
$\pi$ and $K$ exchange diagrams and rigorously including FSI effects 
in a coupled channel approach, a final state conversion is rather 
excluded as dominant origin of the observed $\Sigma^0$ suppression.
While $\Lambda$ production is found to be dominated by kaon exchange 
--- in agreement with the exclusive DISTO spin transfer results at 
higher excess energies~\cite{bal99} --- both $\pi$ and $K$ exchange 
turn out to contribute to the $\Sigma^0$ channel with similar strength.
It is concluded~\cite{gas00}, that a destructive interference of 
$\pi$ and $K$ exchange diagrams might explain the 
close--to--threshold $\Sigma^0$ suppression and a good agreement with 
the COSY--11 total cross section results is obtained after including 
an overall reduction due to the $pp$ initial state 
interaction.

An experimental study of $\Sigma$ production in different isospin 
configurations should provide a crucial test for the above 
interpretation: 
For the reaction $pp \rightarrow n K^+ \Sigma^+$ an opposite 
interference pattern is found as compared to the $p K^+ \Sigma^0$ 
channel, i.e.\ the $n K^+ \Sigma^+$ channel is enhanced for a 
destructive interference of $\pi$ and $K$ exchange.
Measurements close to threshold are planned at the COSY--11 facility 
in future.

Contributions from direct production as well as heavy meson exchanges 
have been neglected so far in these calculations~\cite{gas00} but 
might influence the $\Lambda/\Sigma^0$ production 
ratio~\cite{tsu,kai99}.

Employing both a $\pi$ and $K$ exchange based meson exchange model 
neglecting any interference of the amplitudes and the resonance 
model of~\cite{tsu} the data on the $\Lambda/\Sigma^0$ production 
ratio are described within a factor of two of the experimental error 
bars~\cite{sib00}. 
The same holds for OBE calculations performed by Laget~\cite{laget}, 
in which the relative sign of $\pi$ and $K$ exchange is chosen to 
maximize the cross section. 
It should be noted that the latter approach both reproduces the 
polarization transfer results reported by the DISTO 
collaboration~\cite{bal99} and accurately describes $YN$ invariant 
mass distributions of the inclusive SPES~4 measurements~\cite{sie94}.

Within an effective Lagrangian approach both $\Lambda$ and $\Sigma^0$ 
production are found to be dominated close to threshold by $\pi$ 
exchange followed by an excitation of the $N^*(1650)$ 
resonance~\cite{shy01}, in contrast to the resonance model approach 
considered in~\cite{sib00}, where the influence of the $N^*(1650)$ on 
the $\Sigma^0$ channel has not been taken into account, and it is 
rather the $N^*(1710)$ that determines $\Sigma^0$ production close to 
threshold and both reaction channels at higher excess energies.

Recently, preliminary Dalitz plot distributions obtained at the 
COSY--TOF facility have been presented for the reaction $pp 
\rightarrow p K^+ \Lambda$ at an excess energy of 
$171\,\mbox{MeV}$~\cite{eyr02}.
Qualitatively, the data are reproduced by calculations by 
A.~Sibirtsev considering resonance excitation and effects of 
$p \Lambda$ final state interaction, giving evidence for a dominant 
influence of the $N^*(1650)$ resonance at this excess energy.

\begin{figure}[hbt]
\begin{center}
\epsfig{file=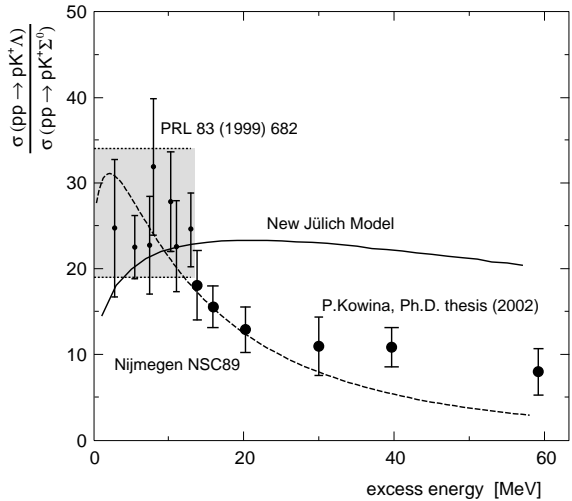,width=0.6\textwidth}
\caption{\label{pkls_ratio} $\Lambda/\Sigma^0$ production ratio in 
proton--proton collisions as a function of the excess energy. 
Experimental data within the shaded area, which corresponds to the 
range given in relation~\eqref{eq_lsratio}, are from~\cite{sew99}, 
data at higher excess energies from~\cite{kow02}. 
Calculations~\cite{gas02} are performed within the J\"ulich meson 
exchange model, assuming a destructive interference of $K$ and $\pi$ 
exchange and employing the microscopic $YN$ interaction models 
Nijmegen NSC89 (dashed line~\cite{mae89}) and the new J\"ulich 
model (solid line~\cite{hai01a}), respectively.}
\end{center}
\end{figure}

Measurements on the $\Lambda/\Sigma^0$ production ratio in 
proton--proton collisions have been extended up to excess energies of 
$\mbox{Q} = 60\,\mbox{MeV}$ at the COSY--11 
installation~\cite{kow02}.
In comparison to the experimental data, in figure~\ref{pkls_ratio} 
calculations are included obtained within the approach 
of~\cite{gas00} assuming a destructive interference of $\pi$ and $K$ 
exchange with different choices of the microscopic hyperon nucleon 
model to describe the interaction in the final state~\cite{gas02}.

As emphasized in~\cite{gas00}, the result depends on the details --- 
especially the off--shell properties --- of the hyperon--nucleon 
interaction employed, although the actual choice does not alter the 
general result in~\cite{gas00} of only a destructive interference of 
$\pi$ and $K$ exchange explaining the experimentally observed 
suppression of the $\Sigma^0$ signal close to threshold.
At the present stage both the good agreement found for J\"ulich 
model~A~\cite{hol89} with the close--to--threshold 
result~\eqref{eq_lsratio} and for the Nijmegen model (dashed line in 
fig.~\ref{pkls_ratio}) with the energy dependence of the cross 
section ratio should rather be regarded as accidental.
In the latter case an SU(2) breaking in the ${}^3\mbox{S}_1$ 
$\Sigma N$ channel had to be introduced~\cite{mae89}.
Consequently, the relation between the $\Sigma^0 p$ amplitude and the 
$\Sigma^+ p$ and $\Sigma^- p$ channels becomes ambiguous.  
Only one of the choices leads to the good agreement with the data, 
whereas the other one results in a completely different 
prediction~\cite{hai02}.

Calculations using the new J\"ulich model (solid line in 
fig.~\ref{pkls_ratio}~\cite{hai01a}) do not reproduce the tendency of 
the experimental data.
It is suggested in~\cite{gas02} that neglecting the energy dependence 
of the elementary amplitudes and assuming S--waves in the final state 
might no longer be justified beyond excess energies of 
$20\,\mbox{MeV}$.

\begin{figure}[hbt]
\begin{center}
\parbox{0.56\textwidth}
  {\epsfig{file=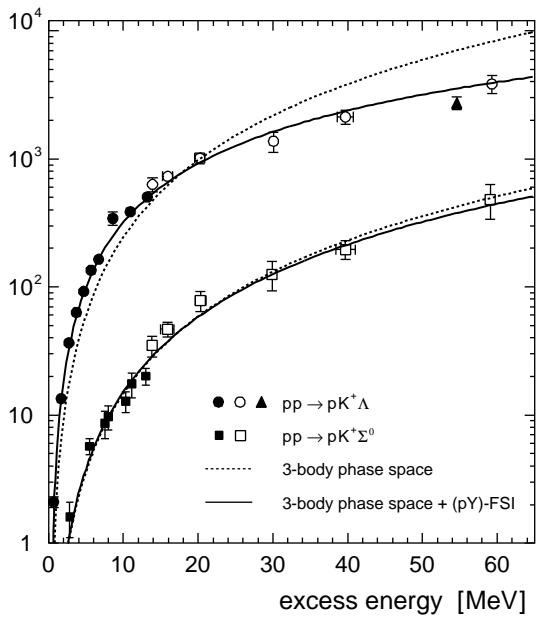,width=0.56\textwidth}} \hfill
\parbox{0.41\textwidth}
  {\caption{\label{pkls_fit} Total cross sections of the reactions $pp 
   \rightarrow p K^+ \Lambda/\Sigma^0$ (solid circles and squares 
   from~\cite{bal98,sew99}, solid triangle from~\cite{bil98}, open 
   symbols from~\cite{kow02}) as functions of the excess energy. 
   Dotted and solid lines denote fits of the energy dependence 
   assuming a pure phase space behaviour and a phase space dependence 
   modified by the proton hyperon FSI parameterized according 
   to~\cite{fae97}. In addition to the absolute normalization, the 
   latter allows to vary low energy scattering parameters via the 
   energy of a nearby virtual bound state.}}
\end{center}
\end{figure}

Total cross sections for the reactions $pp \rightarrow 
p K^+ \Lambda/\Sigma^0$ obtained at the COSY--11 (circles and 
squares~\cite{bal98,sew99,kow02}) and COSY--TOF 
(triangle~\cite{bil98}) facilities up to excess energies of $\mbox{Q} 
= 60\,\mbox{MeV}$ are shown in figure~\ref{pkls_fit}. 

Obviously, in case of $\Lambda$ production, the energy dependence of 
the total cross section is much better described by a phase space 
behaviour modified by the $p \Lambda$ final state interaction than by 
pure phase space.
However, unlike the findings of~\cite{sew99} based on the at that 
time available data up to excess energies of $\mbox{Q} = 
13\,\mbox{MeV}$, in the energy range up to $\mbox{Q} = 
60\,\mbox{MeV}$ $\Sigma^0$ production is equally well described 
neglecting any FSI effect.

Presently, the origin of this qualitatively different behaviour of 
the hyperon production channels is not understood:
Proton--hyperon final state interactions might be less important in 
case of $\Sigma^0$ production compared to $\Lambda$ production as 
already concluded in~\cite{gas99}.
On the other hand, a fit of the energy dependence considering phase 
space and final state interaction effects implies the dominance of 
S--wave production and energy independent reaction dynamics, which 
might no longer be justified as mentioned above~\cite{gas02}.
Within the statistics of the present experiment, P--wave 
contributions can be neither ruled out nor confirmed at higher excess 
energies for $\Sigma^0$ production.
Consequently, high statistics $\Sigma^0$ data would be needed in 
future to study the influence of higher partial waves experimentally.

Qualitatively, a dominant production via resonance excitation --- as 
investigated within the resonance model approach~\cite{tsu} --- might 
provide a mechanism leading to different partial wave contributions, 
if $\Lambda$ and $\Sigma^0$ production close to threshold were 
dominated by the $\mbox{S}_{11}$ $N^*(1650)$ and $\mbox{P}_{11}$ 
$N^*(1710)$ resonances, respectively.

\section{Elementary Antikaon Production}
Studies on the reaction $pp \rightarrow pp K^+ K^-$ close to 
threshold have been motivated by the continuing discussion on the 
nature of the scalar resonances $f_0(980)$ and $a_0(980)$~\cite{f0a0}.
Within the J\"ulich meson exchange model for $\pi \pi$ and $\pi \eta$ 
scattering the $K \overline{K}$ interaction dominated by vector meson 
exchange gives rise to a bound state in the isoscalar sector 
identified with the $f_0(980)$~\cite{kre97}.
Both shape and absolute scale of $\pi \pi \rightarrow K \overline{K}$ 
transitions crucially depend on the strength of the $K \overline{K}$ 
interaction, which in turn is a prerequisite of a $K \overline{K}$ 
molecule interpretation of the $f_0(980)$.
Similar effects might be expected for the elementary kaon--antikaon 
production in proton--proton scattering, and first results of 
exploratory microscopic calculations have recently been 
presented~\cite{hai01b}.

\subsection{Experimental Results}

A first total cross section value for the elementary antikaon 
production below the $\Phi$ threshold in proton--proton scattering has 
been extracted from exclusive data taken at the COSY--11 installation 
at an excess energy of $\mbox{Q} = 17\,\mbox{MeV}$ with 
\begin{equation}
\label{eq_kkcross}
\sigma_{pp \rightarrow pp K^+ K^-} 
  \left(\mbox{Q} = 17\,\mbox{MeV}\right) = 
  1.80 \pm 0.27^{+0.28}_{-0.35}\,\mbox{nb},
\end{equation}
including statistical and systematical errors, 
respectively~\cite{que01}.

\begin{figure}[hbt]
\begin{center}
\epsfig{file=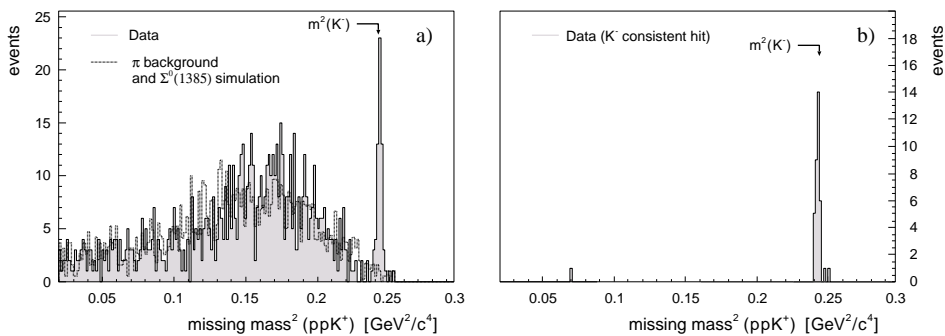,width=\textwidth}
\caption{\label{ppkk_result} Missing mass squared with respect to an 
identified $(pp K^+)$ subsystem at an excess energy of 
$17\,\mbox{MeV}$ above the $pp \rightarrow pp K^+ K^-$ threshold 
a) without and b) with $K^-$ detection. In a) the distribution at 
lower missing mass values is reproduced considering the possible 
misidentification of pions as kaons as well as the excitation of 
hyperon resonances (black solid line).}
\end{center}
\end{figure}

The experimental technique is based on the measurement of the complete 
four--momenta of positively charged ejectiles.
Figure~\ref{ppkk_result}a) shows the missing mass squared with respect 
to an identified $(pp K^+)$ subsystem:
A sharp peak at the charged kaon mass corresponding to a resolution 
of $\approx 2\,\mbox{MeV}/\mbox{c}^2$ (FWHM) is clearly separated 
from a broad distribution in the region of lower missing masses.
The latter can be explained by misidentifying pions from $pp 
\rightarrow pp \pi^+ X$ events as kaons, where $X$ denotes a system 
of undetected particles.
In addition, the excitation of hyperon resonances via $pp \rightarrow 
p K^+ \Lambda(1405)/\Sigma(1385)$ may contribute, with one of the 
identified protons originating from the hyperon resonance decay, 
shifting the missing mass with respect to the $(ppK^+)$ subsystem to 
lower values (as discussed in detail in~\cite{que01}).
Considering both effects, the broad distribution is well reproduced 
as indicated in figure~\ref{ppkk_result}a).

Requiring furthermore a $K^-$ consistent hit in the dedicated 
negative particle detection system of the COSY--11 
facility~\cite{bra96}, the identification of the four particle final 
state becomes (almost) completely free of background, as demonstrated 
in figure~\ref{ppkk_result}b).
The reduction in counting rate for the $K^-$ signal is due to 
acceptance and decay losses and in excellent agreement with 
expectations from Monte Carlo simulations.

However, the available statistics of $K^+ K^-$ events extracted at 
the excess energy of $\mbox{Q} = 17\,\mbox{MeV}$ is not sufficient to 
distinguish between a non--resonant $K^+ K^-$ production and resonant 
production via the scalar resonances $f_0(980)$ and $a_0(980)$ from 
differential observables, e.g.\ the $pp$ missing mass 
distribution~\cite{kkres}.

\subsection{Energy Dependence of the Total Cross Section}
The total $\eta$, $\omega$ and $\eta^\prime$ production cross sections 
show very similar dependences on the excess energy:
At excess energies $100 \le \mbox{Q} \le 1000\,\mbox{MeV}$ the 
energy dependence of the total cross section is dominated by 
three--body phase space ($\sigma \propto \mbox{Q}^2$).
The deviation from a $\mbox{Q}^2$ dependence below $100\,\mbox{MeV}$ 
arises from the interaction between the final state protons and 
possibly between the final state proton and meson, the latter clearly 
observed in case of the $pp \rightarrow pp \eta$ reaction. 

\begin{figure}[hbt]
\parbox{0.49\textwidth}{\epsfig{file=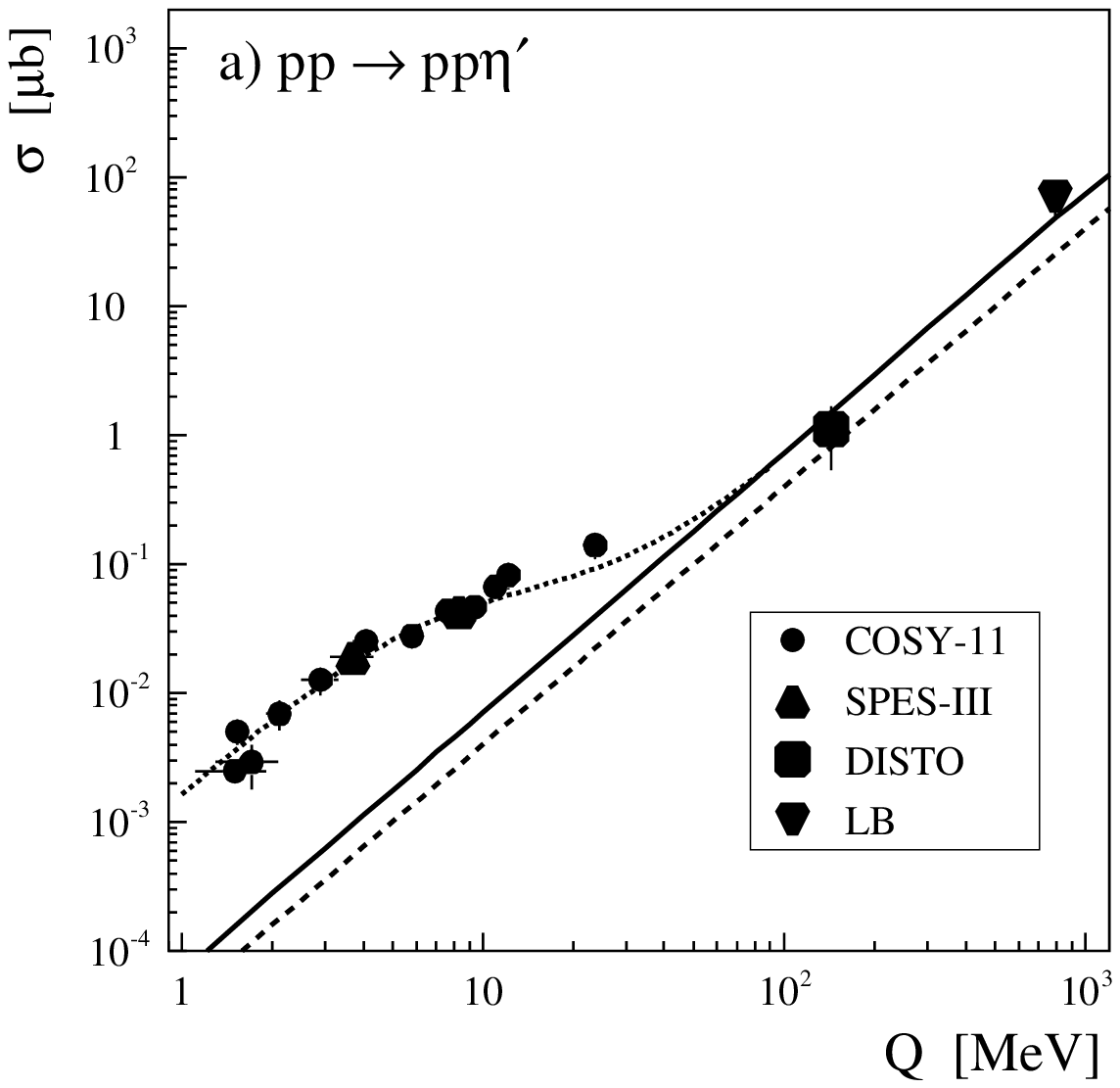,width=0.49\textwidth}}
\hfill
\parbox{0.49\textwidth}{\epsfig{file=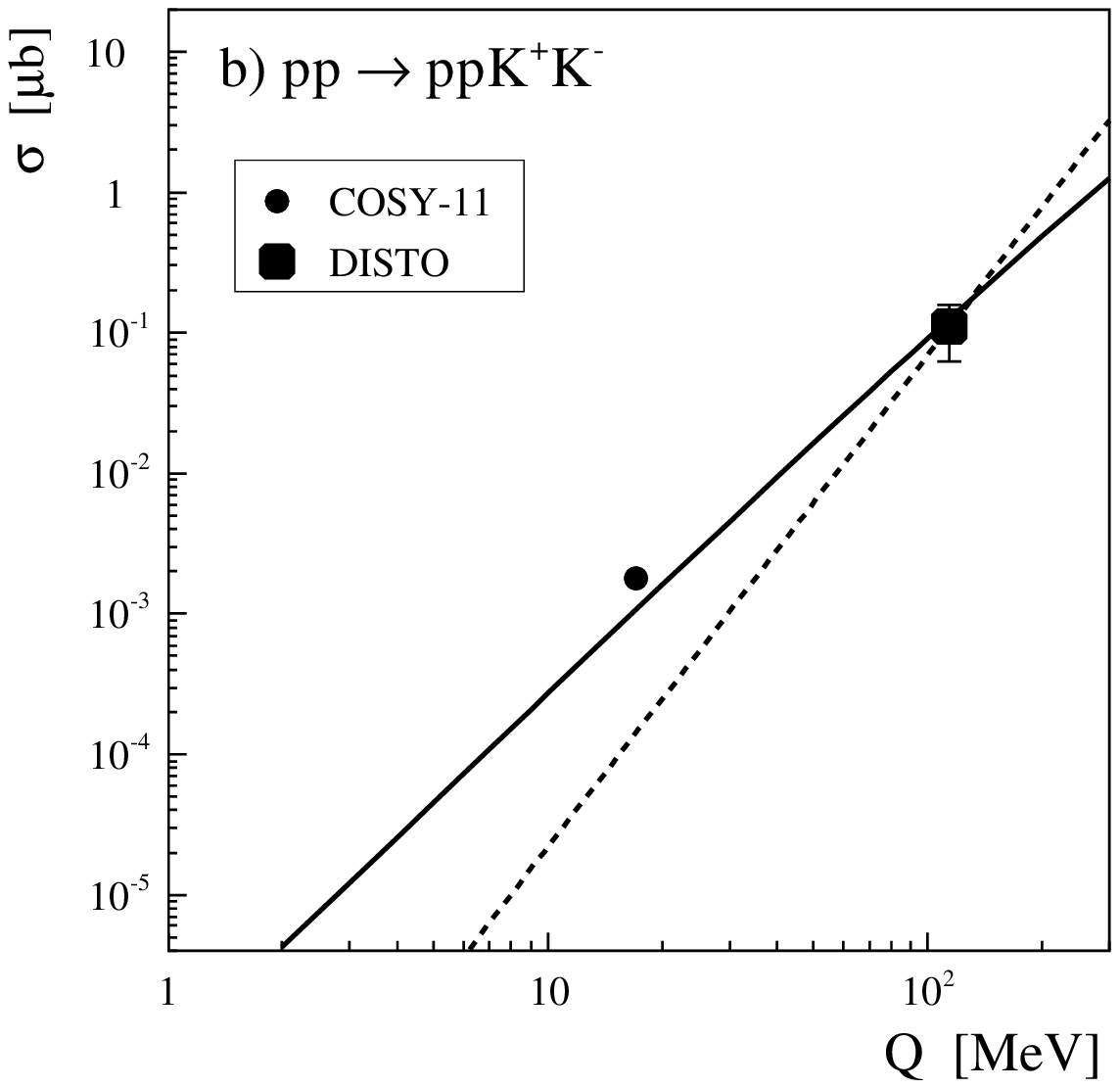,width=0.49\textwidth}}
\caption{\label{kminus1}a) The $pp \rightarrow pp \eta^\prime$ cross 
section as a function of excess energy. The data are 
from~\cite{bal88,etap}, the dashed line shows the phase space 
$\mbox{Q}^2$--dependence, the solid line indicates calculations 
without FSI~\cite{sibbaru}, the dotted line shows the 
parameterization of the $p p$ FSI. 
b) The $pp \rightarrow pp K^+ K^-$ cross section. The data are 
from~\cite{que01,bal01}, the solid line shows the calculations 
of~\cite{sib97}, the dashed line indicates the phase space 
$\mbox{Q}^{7/2}$--dependence.}
\end{figure}

These features are well illustrated by the data~\cite{bal88,etap} on 
the reaction $pp \rightarrow pp \eta^\prime$, where the possible 
effect due to the $p \eta^\prime$ FSI is expected to be almost 
negligible.
Figure~\ref{kminus1}a) shows the data available for the $pp 
\rightarrow pp \eta^\prime$ cross section as a function of excess 
energy Q.
The phase space dependence (dashed line) apart from the normalization 
constant reproduces the data at $\mbox{Q} \ge 100\,\mbox{MeV}$.
Calculations~\cite{sibbaru} obtained within one--boson exchange 
models neglecting the $pp$ FSI explicitly follow the phase space 
dependence (solid line), while the effect of the $pp$ FSI is indicated 
by the dotted line.

Data on the $pp \rightarrow pp K^+ K^-$ total cross section obtained 
at the COSY--11~\cite{que01} and DISTO~\cite{bal01} facilities below 
and above the $\Phi$ production threshold, respectively, are in 
reasonable agreement with one--boson exchange 
calculations~\cite{sib97} without FSI effects (solid line in 
figure~\ref{kminus1}b)).
Contrary to $\eta$, $\omega$ and $\eta^\prime$ production the 
calculations for $K^+ K^-$ production differ significantly from the 
four--body phase space behaviour (dashed line).
The latter effect can be understood in terms of the energy dependence 
of the elementary scattering amplitudes which determine the energy 
dependence of the $pp \rightarrow pp \eta^\prime$ and $pp \rightarrow 
pp K^+ K^-$ total cross sections:
While the $\pi N \rightarrow \eta^\prime N$ amplitudes are almost 
independent of the invariant $\pi N$ energy, 
$K^+ p$ and especially $K^- p$ scattering data exhibit a substantial 
energy dependence~\cite{sib01}.
Comparing the COSY--11 result~\eqref{eq_kkcross} with the 
calculations shown by the solid line in figure~\ref{kminus1}b) --- 
neglecting FSI --- one might detect no room for final state 
interaction effects.
Contrary to this, $\eta$, $\omega$ and $\eta^\prime$ production 
indicate strong FSI imprints at excess energies $\mbox{Q} \le 
100\,\mbox{MeV}$ (fig.~\ref{kminus1}a)).

Presently it is not clear whether the absence of the FSI influence in 
the $pp \rightarrow pp K^+ K^-$ reaction might be explained by a 
partial compensation of the $pp$ and $K^- p$ interaction in the final 
state or by the additional degree of freedom given by the four--body 
final state.
In the latter case FSI effects are expected to be more pronounced at 
energies very close to the $K^+ K^-$ production threshold.
It should be noted, that in the presence of two strongly interacting 
particles in the final state --- $pp$ and $K^- p$ --- a factorization 
in terms of two--body interactions might no longer be valid and one 
would face a four--body problem.
Thus, further measurements provide a unique opportunity to get 
insight into the problem experimentally~\cite{sib01}.

Data taking at excess energies closer to threshold and slightly below 
the $\Phi$ production threshold, i.e.\ at excess energies of 
$10\,\mbox{MeV}$ and $28\,\mbox{MeV}$ with respect to the $K^+ K^-$ 
threshold, has been successfully completed early this year at the 
COSY--11 facility and data analysis is presently in progress.

\section{Acknowledgements}
We gratefully acknowledge valuable discussions with A. Gasparian, 
J. Haidenbauer, C. Hanhart and A. Sibirtsev.
This work was partly supported by the European Community -- Access to 
Research Infrastructure action of the Improving Human Potential 
Programme.

\end{document}